\documentclass[conference]{IEEEtran}

\normalsize
\ifCLASSINFOpdf
\else
\fi

\usepackage{color}
\usepackage{amsfonts}
\usepackage{amssymb}
\usepackage{amsmath}
\usepackage[dvips]{graphicx}
\usepackage{cite}
\usepackage{balance}
\usepackage{psfrag}
\usepackage{textcomp}
\usepackage{float}
\usepackage{esint}

\newtheorem{theorem}{Theorem}

\newtheorem{remark}[theorem]{Remark}

\newtheorem{innercustomgeneric}{\customgenericname}
\providecommand{\customgenericname}{}
\newcommand{\newcustomtheorem}[2]{%
  \newenvironment{#1}[1]
  {%
   \renewcommand\customgenericname{#2}%
   \renewcommand\theinnercustomgeneric{##1}%
   \innercustomgeneric
  }
  {\endinnercustomgeneric}
}

\newcustomtheorem{prop}{Proposition}
\newcustomtheorem{cor}{Corollary}

\title{Analytical Performance Evaluation of Beamforming Under Transceivers Hardware Imperfections}

\author{%
\authorblockN{Alexandros--Apostolos A. Boulogeorgos\authorrefmark{1}, Angeliki~Alexiou\authorrefmark{1}}
\authorblockA{\authorrefmark{1}\footnotesize  Department of Digital Systems, University of Piraeus, Piraeus 18534,
Greece. (E-mails: al.boulogeorgos@ieee.org, alexiou@unipi.gr).}
}

\author{
Alexandros--Apostolos A. Boulogeorgos, and~Angeliki~Alexiou

\\
\begin{normalsize} 
Department of Digital Systems, University of Piraeus, Piraeus 18534, Greece.
\end{normalsize}
\\
\begin{normalsize} 
e-mail:  al.boulogeorgos@ieee.org, alexiou@unipi.gr.
\end{normalsize} 
}

\begin{document}

\maketitle	

\begin{abstract}

In this paper, we provide the mathematical framework to evaluate and quantify the performance of wireless  systems, which employ beamforming, in the presence of hardware imperfections at both the basestation and the user equipment. 
In more detail, by taking a macroscopic view of the joint impact of hardware imperfections, we introduce a general model that accounts for transceiver impairments in beamforming transmissions. 
In order to evaluate their impact, we present novel closed form expressions for the outage probability and upper bounds for the characterization of the system's capacity. 
Our analysis reveals that the level of imperfection can significantly constraint the ergodic capacity. 
Therefore, it is important to take them into account when evaluating and designing  beamforming systems.

\end{abstract}

\section{Introduction}\label{S:Intro}

The increased data rates demand, in the beyond fifth generation (5G) wireless systems, as well as the fact that the spectral efficiency of the microwave links is approaching its fundamental limits have motivated the exploitation of higher frequency bands that offer abundance of communication bandwidth~\cite{A:THz_Technologies_to_deliver_Optical_network_quality_of_experience_in_Wireless_Systems_beyond_5G,
WP:Wireless_Thz_system_architecture_for_networks_beyond_5G,
A:LC_CR_vs_SS}.  
Towards this end, higher unlicensed bands, such as millimeter wave (mmW) and terahertz (THz), for wireless communications have been recognized as attractive canditates~\cite{A:Short_range_Ultra_broadband_THz_Communications_Concepts_and_perspectives, A:THz_Band_New_frontier_for_wireless_communications}.  

Although these communication bands can benefit from an extreme increase in the bandwidth, it comes at the price of suffering severe path loss, which drastically limits the links range~\cite{C:PerfEvaluation, C:ChannelModelTHz}. 
To overcome this problem, a great amount of research effort has been put on investigating the use of multiple antennas in order to achieve link-level directional beamforming~\cite{A:Indoor_THz_Communications_How_Many_Antenna_Arrays_are_needed,
A:OnMillimeterWaveAndTHzMobileRadioChannelForSmartRailMobility,
A:Analytical_Performance_Assessment_of_THz_Wireless_Systems}. 
However, most of the published papers in the context of  mmW and THz wireless systems make the standard assumption of ideal transceiver~hardware. 

In practice, hardware suffers from several types of impairments, such as phase noise, in-phase (I) and quadrature (Q) imbalance, and amplifier nonlinearities (see \cite{B:Schenk-book} and reference therein). 
The impact of hardware impairments  on various types of communication systems was analyzed in~\cite{ Performance_of_MIMO_OFDM_Systems_with_Phase_Noise_at_Transmit_and_Receive_Antennas, 5464254, A:IQI_in_AF_Nakagami_m, RF_impairments_generalized_model, A:A_new_look_at_dual_hop_relaying_performance_limits_with_hw_impairments}. 
In particular, in~\cite{Performance_of_MIMO_OFDM_Systems_with_Phase_Noise_at_Transmit_and_Receive_Antennas}, the authors investigated the impact of phase noise in multiple-input multiple-ouput (MIMO) orthogonal frequency division multiplexing (OFDM) systems, whereas, in~\cite{5464254}, the authors analyzed the performance of MIMO system in the presence of I/Q imbalance. Moreover, in~\cite{ A:IQI_in_AF_Nakagami_m}, the impact of I/Q imbalance in amplify and forward (AF) relaying systems was quantified. Finally, a generalized model that takes into account the joint impact of I/Q imbalance, phase noise and non-linearities was presented in~\cite{RF_impairments_generalized_model}, while, in~\cite{A:A_new_look_at_dual_hop_relaying_performance_limits_with_hw_impairments}, the authors used this model to evaluate the impact of hardware imperfections in dual hop AF relaying systems.
As a general conclusion, hardware impairments have a deleterious impact on the achievable performance, especially in high data rate systems~\cite{B:Schenk-book,A:Energy_detection_spectrum_sensing_under_RF_imperfections, A:OFDM_OR_IQI,6554955,
C:Energy_Detection_under_RF_impairments_for_CR}. 

In spite of the paramount importance of hardware impairments on wireless systems, their detrimental effect has been overlooked in the analysis of  wireless systems that employ beamforming. Motivated by this, the present work is devoted to the quantification of the impact of hardware impairments on the performance of beamforming transmissions. In more detail, the technical contribution of the paper is outlined below:
\begin{itemize}
\item We introduce a general model to account for transceiver impairments in beamforming transmissions. 
Unlike~\cite{A:Widely_linear_beamforming_and_RF_impairment_suppression_in_massive_antenna_arrays,7112184}, which study the effect of a single type imperfection, in this work, we take a macroscopic view and examine the joint impact of hardware impairments. 
\item After deriving the instantaneous signal-to-noise-plus-distortion-ratio (SNDR), we present a novel closed-form expressions for the exact outage probability of the system. This expression enables the quantification of the impact of hardware imperfection of the performance of the systems. Additionally, they can be used to select the transmission data rate and the number of antennas that are required to satisfy an outage probability specification. Moreover, we present simple closed-form expressions for the outage probability that corresponds to the special case in which all radio frequency (RF) chains suffer from the same level of imperfections.  
\item Moreover, in order to characterize the ergodic capacity of the beamforming channel, we derive a simple close-form expression for its upper bound. 
\item Finally, we provide simple expressions for the capacity ceilings, when the signal-to-noise ratio (SNR) and or the number of transmit antennas tend to infinity.
\end{itemize}

The remainder of the paper is organized as follows: Section~\ref{S:SM} presents the system model, while Section~\ref{S:OP} is devoted to the derivation of the analytic expression for the outage probability. Respective numerical results and discussions are provided in Section~\ref{S:Results}, whereas closing remarks are presented in Section~\ref{S:Conclusions}.

\textit{Notations}: Unless otherwise stated, lower and upper case  bold letters denote vectors and matrices, respectively. The matrix $\mathbf{0}_{N\times M}$ stand for the all zero $N\times M$ matrix. The operator $\mathrm{diag}\left(\mathbf{x}\right)$ returns a matrix with the elements of the vector $\mathbf{x}$ on the leading diagonal, and $0$ elsewhere. Also, $(\cdot)^*$ and $(\cdot)^t$ denote the matrix complex-conjugate  and  transpose  operations,  respectively. 
The operator $||\mathbf{x}||$ represents the norm of the vector $\mathbf{x}$, while $|x|$ is the absolute time of the variable $x$. 
$\mathbb{E}[\mathcal{X}]$ stands for the expected value of the random variable $\mathcal{X}$, while $\exp(x)$ and $\log_2(x)$ return the exponential and the the base-2 logarithm of $x$, respectively.  
Finally, the lower incomplete Gamma functionis represented as $\gamma\left(\cdot,\cdot\right)$~\cite[Eq. (8.350/1)]{B:Gra_Ryz_Book}, while the Gamma function is denoted by $\Gamma\left(\cdot\right)$~\cite[Eq. (8.310)]{B:Gra_Ryz_Book}. 

\section{System and Signal Model}\label{S:SM}

\begin{figure}
\psfrag{A}[][][1]{$1$}%
\psfrag{B}[][][1]{$2$}%
\psfrag{C}[][][1]{$N_t$}%
\psfrag{D}[][][1]{BS}%
\psfrag{E}[][][1]{UE}%
\centering\includegraphics[width=0.6\linewidth,trim=0 0 0 0,clip=false]{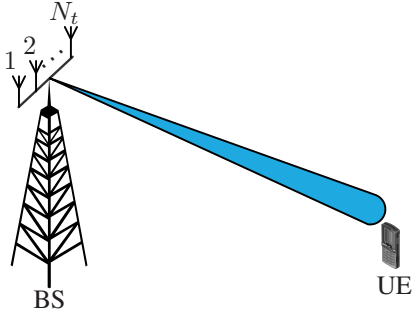}
\caption{System model.}\label{fig:SM}
\end{figure}

As presented in Fig.~\ref{fig:SM}, 
we consider a downlink scenario of a classical wireless system, in which the basestation (BS) is equipped with $N_t$ antennas, while the user equipment (UE) has a single antenna. The BS is assumed to perform digital beamforming, i.e., each antenna is driven by an individual RF chain. Moreover, we assume perfect channel state information (CSI) at the BS\footnote{The  case  of  imperfect  CSI  due  to  hardware imperfections  and  outdated  estimation  will  not be  investigated in this paper,  due  to  space  limitations.  However,  the  performance  results presented  here are upper bounds  to those when imperfect  CSI is assumed. Moreover, note that in the case of using a time division duplexing scheme, since the channel reciprocity property is valid, the BS can estimate the downlink channel~\cite{Zetterberg:2011:EIT:1928509.1972710,4395422}.}. 

At the BS, the information symbol, $s$, is pre-coded using the $N_t\times 1$ beamforming vector, $\mathbf{p}$. We set $||\mathbf{p}||=1$, to reflect the power constraint at the BS. Note that the maximum signal to noise ratio (SNR) is achieved, if $\mathbf{p}$ is proportional to $\mathbf{h}^{*}$,~i.e.,
\begin{align}
\mathbf{p} = \frac{\mathbf{h}^{*}}{||\mathbf{h}^{*}||},
\end{align}     
where $\mathbf{h}$ stand for the baseband equivalent channel. 

The transmitted signal, $\mathbf{x}=\mathbf{p} s$, is conveyed over the flat fading wireless channel, $\mathbf{h}$, with additive noise $n\in\mathbb{C}$. Note that this channel can, for instance, be one of the subcarriers in a multi-carrier system. By assuming ideal radio frequency (RF) front end at both the BS and  UE, the received signal can be modeled~as 
\begin{align}
y = \mathbf{h}^t \mathbf{x} + n,
\label{Eq:CovSignalModel}
\end{align}
where $\mathbf{h}^t$ and ${n}$ are statistically independent. Note that due to the relatively small bandwidth of the subcarrier, $n$ can be approximated as a zero mean complex Gaussian process with single-sided power spectrum  densities (PSD) of $N_0$~\cite{5995306}. Moreover, the channel vector 
\begin{align}
 \mathbf{h}=&\left[|h_1|\exp(-j2\pi\phi_1), |h_2| \exp(-j2\pi\phi_2), \cdots, 
 \right. \nonumber \\ &\hspace{+3.5cm}\left.|h_{N_t}|\exp(-j2\pi\phi_{N_t}) \right]^t,
 \end{align}
 where $|h_i|$ and $\phi_i$, $i\in[1, N_t]$, denote the amplitude and phase of the channel of the $i-$th BS antenna and the UE. Finally, it is assumed that $|h_i|$ follows Rayleigh\footnote{Note that this approach has been proven to be an appropriate model for indoor THz communications ~\cite{A:Indoor_THz_Communications_How_Many_Antenna_Arrays_are_needed,7582545,A:StatisticalModelingAndSimulationOfShortRangeD2D_CommunicationChannelsAtSub_THzFrequency}.}.

The conventional received signal model, described by~\eqref{Eq:CovSignalModel}, cannot capture the impact of hardware imperfections. Due to these imperfections,  each RF chain, $i$ with $i\in[1, N_t]$, at the BS causes a mismatch between the intended transmitted signal, $x_i$, and what is actually generated and emitted. As a consequence, the actual transmitted signal can be modeled~as
\begin{align}
\tilde{\mathbf{x}}= \mathbf{x} + \mathbf{W}_b \mathbf{p}, 
\end{align}   
where $ \mathbf{W}_b=\mathrm{diag}\left([w_{b,1}, w_{b,2}, \cdots, w_{b,N}]\right)$ represent the distortion noise from impairments at the BS. Similarly, at the UE, the baseband equivalent received signal can be expressed~as
\begin{align}
\tilde{y} = \mathbf{h}^t \left(\mathbf{x}+ \mathbf{W}_b \mathbf{p}\right) + w_{u} + n,
\label{Eq:received_signal_impaired}
\end{align} 
where $w_u$ stands for the distortion noise from the impairments at the UE. The distortion noises are defined as~\cite{B:Schenk-book,A:A_new_look_at_dual_hop_relaying_performance_limits_with_hw_impairments,A:Energy_detection_spectrum_sensing_under_RF_imperfections}
\begin{align}
\mathbf{w}_b&\sim\mathcal{CN}\left(\mathbf{0}_{N_t\times N_t}, \mathrm{diag}\left(\left[k_{b,1}^2, k_{b,2}^2, \cdots, k_{b,N}^2\right]\right) P_s\right), \\
w_u&\sim\mathcal{CN}\left(0, k_u^2 P_s ||\mathbf{h}||^2\right),
\label{Eq:s_w_u}
\end{align} 
where $k_{b,i}$ ($i=1, 2, \cdots, N$) and $k_u$ are non-negative design parameters that characterize the level of imperfection in the BS and UE hardware, respectively, whereas $P_s$ stands for the average received signal power.  This model has been validated by several theoretical and experimental results, see e.g., \cite{MIMO_transmission_with_residual_transmit_RF_impairments, B:wenk2010mimo, C:Assessing_and_Modeling_the_effect_of_RF_impairments_on_ULTRA_LTE_Uplink_performance} and references therein. 
Note that for $k_{b,1}=k_{b,2}=\cdots=k_{b,N}=k_u=0$, i.e., in the case of ideal RF chains at both the BS and UE, \eqref{Eq:received_signal_impaired} is simplified into~\eqref{Eq:CovSignalModel}.

\section{Performance Analysis}\label{S:OP}
In this section, after we derive the SNDR, we evaluate the system's outage probability and we characterize the ergodic~capacity. 

\subsection{SNDR}
Based on~\eqref{Eq:received_signal_impaired}-\eqref{Eq:s_w_u} and after some basic algebraic manipulations, the SNDR can be expressed as
\begin{align}
\gamma=\frac{||\mathbf{h}||^2 P_s}{\sum_{i=1}^N (k_{b,i}^2+k_u^2)|h_{0i}|^2 P_s + N_0}.
\label{Eq:SDNR}
\end{align} 
From~\eqref{Eq:SDNR}, it is evident that as the number of antennas increases the diversity gain increases together with the level of interference, due to the hardware imperfections. 
Moreover, a transmission power increase leads to a linear increase of the interference level. 
Finally, we observe that as the level of imperfection increases, i.e., as $k_{b,i}^2+k_u^2$, $i=1, 2, \cdots, N_t$, the SNDR decreases. 
Note that according to the third generation partnership project (3GPP) long term evolution advanced (LTE-A), the parameters $k_{b,i}$ and $k_u$ are in the range of $[0.08, 0.175]$~\cite{Holma:2011:LUE:2028649}.

\subsection{Outage probability}
In order to quantify the impact of hardware imperfections on the system's performance, as well as to provide design criteria  for wireless systems that employ digital beamforming schemes, we derive the outage probability. The following proposition returns the outage probability.

\begin{prop}{1}\label{prop1} The outage probability can be obtained~as
\begin{align}
P_{o}(r_{\mathrm{th}}) = \sum_{i=1}^{N_t} \Xi(i, 1, b_i) \left(1 - \exp\left({-\frac{2^{r_{\mathrm{th}}}-1}{\tilde{\gamma} b_i}}\right)\right),
\label{Eq:OP}
\end{align} 
where  $r_{\mathrm{th}}$ is  the  minimum  allowable  (threshold)  rate, $\Xi(i, 1, b_i)$ is defined in \cite[Eqs. (8) and (9)]{A:Karagiannidis-2006-ID448},
\begin{align}
b_{i} = 1-\left(k_{b,i}^2 + k_u^2\right) \left(2^{r_{\mathrm{th}}}-1\right),
\end{align}
and
\begin{align}
\tilde{\gamma}=\frac{P_s}{N_0}.
\end{align}
\end{prop}
\begin{IEEEproof}
Please refer to the Appendix A.
\end{IEEEproof}
Note that $\Xi\left(\cdot,1, b_i\right)$ depends  on the level of hardware imperfections and $r_{\mathrm{th}}$. In other words, in contrast  with the ideal RF chains scenario, the outage probability does not only depends on the SNR and the $r_{\mathrm{th}}$, but also on the values of $k_{b,i}$, $i=1, 2, \cdots, N_t$ and~$k_u$. 

The following remark presents a design criterion that should be taken into consideration when designing a digital beamforming scheme. 

\begin{remark} From~\eqref{Eq:X}, it is observed that a link between the $i-$th antenna of the BS and the UE, in which $\left(k_{b,i}^2 + k_u^2\right)\left(2^{r_{\mathrm{th}}}-1\right)>1$, has a negative impact on the performance of the beamforming scheme.  
\end{remark}

\subsubsection*{Special case} In the special case in which $k_{b,1}=k_{b,2}=\cdots=k_{b,N_t}=k_b$, $b_1=b_2=\cdots=b_{N_t}=b$ and the random variable $\mathcal{X}$, defined in Appendix B~\eqref{Eq:X}, follows chi-square distribution, which CDF given~by
\begin{align}
F_{\mathcal{X}}^{\mathrm{sc}}(x) = \frac{\gamma\left(N_t, \frac{x}{b}\right)}{\Gamma{(N_t)}},
\end{align}
which, by taking into account that $N_t$ is an integer, can be simplified as
\begin{align}
F_{\mathcal{X}}^{\mathrm{sc}}(x) = 1 - \sum_{k=0}^{N_t-1}\frac{x^k}{b^k k!}\exp\left(-\frac{x}{b}\right).
\end{align}
Hence, in this case, the outage probability can be evaluated~as
\begin{align}
P_{o}^{\mathrm{sc}}=F_{\mathcal{X}}^{\mathrm{sc}}\left( \frac{\gamma_{\mathrm{th}}}{\tilde{\gamma}}\right).
\label{Eq:PoSC}
\end{align}
or, equivalently
\begin{align}
P_{o}^{\mathrm{sc}}=1 - \sum_{k=0}^{N_t-1}\frac{\gamma_{\mathrm{th}}^k}{\tilde{\gamma}^k b^k k!}\exp\left(-\frac{\gamma_{\mathrm{th}}}{\tilde{\gamma} b}\right)
\end{align}

\subsection{Ergodic capacity}
To characterize the ergodic capacity of the beamforming channel, an upper bound is derived by the following~proposition.

\begin{prop}{2}\label{prop2}
The ergodic capacity, $C$, (in bits/channel use) with beamforming and non-ideal hardware is upper bounded~as
\begin{align}
C\leq \log_{2}\left( 1 + \frac{ N_t}{\sum_{i=1}^{N_t}{\Xi(i,1,c_i)}{c_i} + \frac{1}{\tilde{\gamma}}}\right),
\label{Eq:C_bound}
\end{align}
where $c_i$ can be obtained~as
\begin{align}
c_i = k_{b,i}^2 + k_u^2, 
\end{align}
with $i\in[1, N_t]$. 
\end{prop}

\begin{IEEEproof}
Please refer to the Appendix B.
\end{IEEEproof}

\begin{cor}{1}[Capacity ceiling]\label{CorollaryCeiling} As the SNR tends to infinity, the ergodic capacity is constraint~by
\begin{align}
\lim_{\tilde{\gamma}\to\infty}C = \log_2\left(1+\frac{N_t}{ \sum_{i=1}^{N_t}{\Xi(i,1, c_i)}{c_i}}\right).
\end{align} 
\end{cor}
\begin{IEEEproof}
Based on~\eqref{Eq:C_bound},   the capacity ceiling can be expressed~as
\begin{align}
\lim_{\tilde{\gamma}\to\infty}C &\leq \lim_{\tilde{\gamma}\to\infty}\left(\log_{2}\left( 1 + \frac{ N_t}{\sum_{i=1}^{N_t}{\Xi(i,1,c_i)}{c_i} + \frac{1}{\tilde{\gamma}}}\right)\right) \nonumber \\
&= \log_{2}\left( 1 + \frac{ N_t}{\sum_{i=1}^{N_t}{\Xi(i,1,c_i)}{c_i} }\right).
\end{align}
This concludes the proof. 
\end{IEEEproof}

\subsubsection*{Special case} In the special case in which $k_{b,1}=k_{b,2}=\cdots=k_{b,N_t}=k_b$, the random variable $\mathcal{Y}$, defined in~\eqref{Eq:Y}, follows chi-square distribution, with $N_{t}$ degrees of freedom; hence, $\mathbb{E}\left[\mathcal{Y}\right]=N_t\left(k_{b}^2+k_u^2\right)$, and the ergodic capacity is upper bounded~as
\begin{align}
C^{\mathrm{sc}}\leq\log_2\left(1 + \frac{N_t}{N_t\left(k_{b}^2+k_u^2\right)+\frac{1}{\tilde{\gamma}}}\right).
\label{Eq:CapacityBoundSpecialCase}
\end{align}

From~\eqref{Eq:CapacityBoundSpecialCase}, it is evident that in the high SNR regime, the ergodic capacity satisfies
\begin{align}
\lim_{\tilde{\gamma}\to\infty}C^{\mathrm{sc}} = \log_2\left(1 + \frac{1}{k_{b}^2+k_u^2}\right).
\label{Eq:limit1}
\end{align}
Moreover, for a given $\tilde{\gamma}$, as the number of antennas increases, the capacity tends to
\begin{align}
\lim_{N_t\to\infty}C^{\mathrm{sc}} = \log_2\left(1 + \frac{1}{k_{b}^2+k_u^2}\right).
\label{Eq:limit2}
\end{align} 
From~\eqref{Eq:limit1}, we observe that in the high SNR regime, the performance of the system is independent from the number of antennas at the BS and   are fully determined by the level of imperfections. 
Finally, \eqref{Eq:limit2} indicates the detrimental effect of hardware imperfection in massive MIMO systems that employ digital beamforming.

\section{Results \& Discussion}\label{S:Results}
In this section, we validate the theoretical analysis with Monte-Carlo simulations. Furthermore, it is important to note that, unless otherwise is stated, in the following figures, the numerical results are shown with continuous lines, while markers are employed to illustrate the simulation results. 

\begin{figure}
\centering\includegraphics[width=0.9\linewidth,trim=0 0 0 0,clip=false]{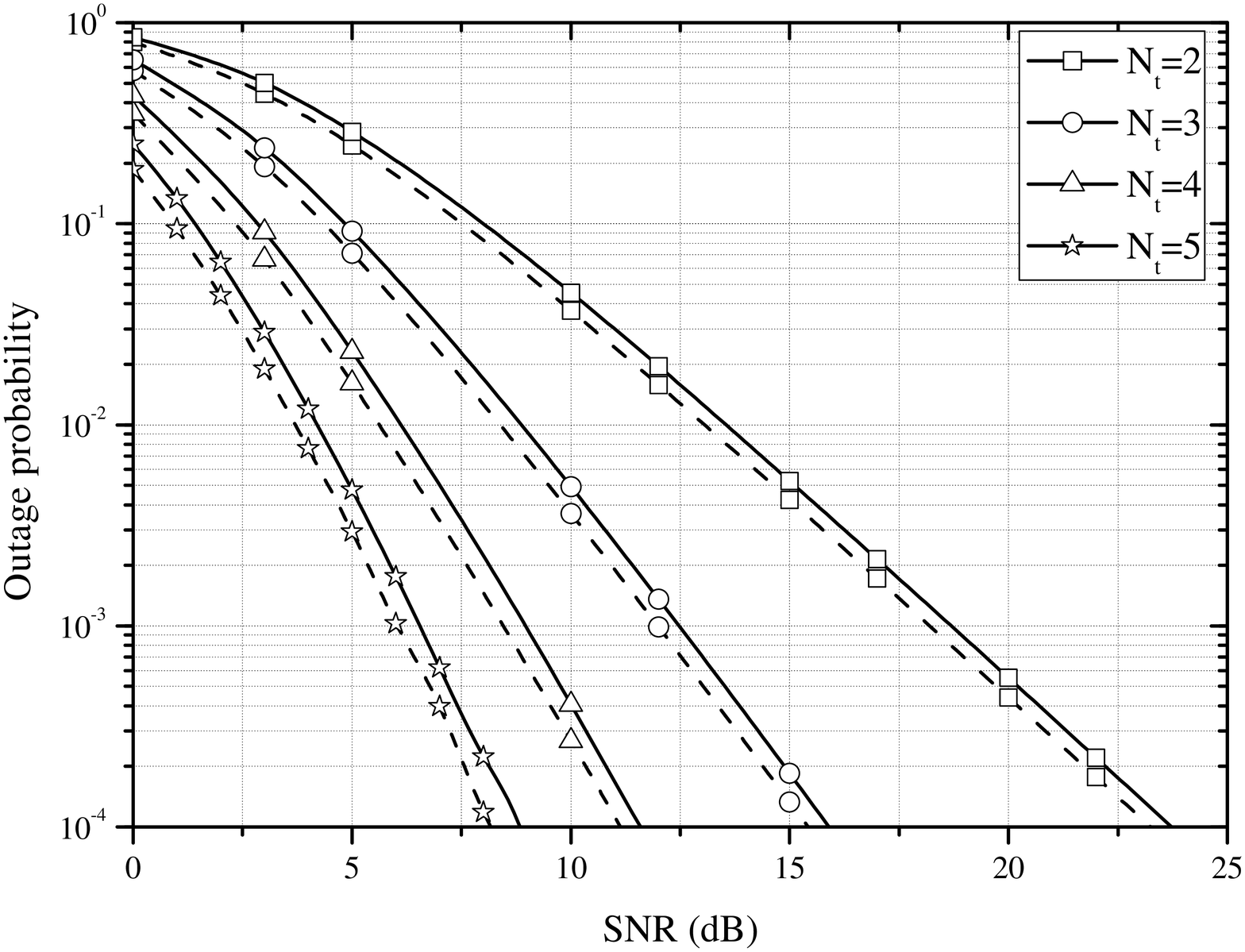}
\caption{Outage probability as a function of the SNR for different values of~$N_t$, assuming ideal (dashed lines) and non-ideal (continious lines) RF chains with $k_{b,i}=0.16$, $i\in[1, N_t]$, $k_u=0.1$, and $r_{\mathrm{th}}=2$.}\label{fig:OP_vs_SNR}
\end{figure}

To this end, Fig.~\ref{fig:OP_vs_SNR} illustrates the outage probability versus the SNR for different values of $N_t$, assuming that $k_{b,i}=0.16$, $i\in[1, N_t]$, $k_u=0.1$, and $r_{\mathrm{th}}=2$ (continuous lines). Moreover, in this figure, as a benchmark, we include the corresponding curves when the RF chains at both the BS and UE are considered ideal (dashed lines). As expected, for a given $N_t$, as the SNR increases, the outage probability decreases, whereas, for a given SNR, as $N_t$ increases, the outage probability decreases. Moreover, the impact of hardware imperfections become more severe as $N_t$ increases. For instance, for SNR equals $10$ dB and $N_t=2$, the performance loss, due to hardware imperfections is $22.44\%$, whereas for the same SNR and $N_t=4$, the performance loss equals $51.94\%$. This indicates the importance of taking into consideration the impact of hardware imperfection, when we select the number of operational RF front-ends.  

\begin{figure}
\centering\includegraphics[width=0.9\linewidth,trim=0 0 0 0,clip=false]{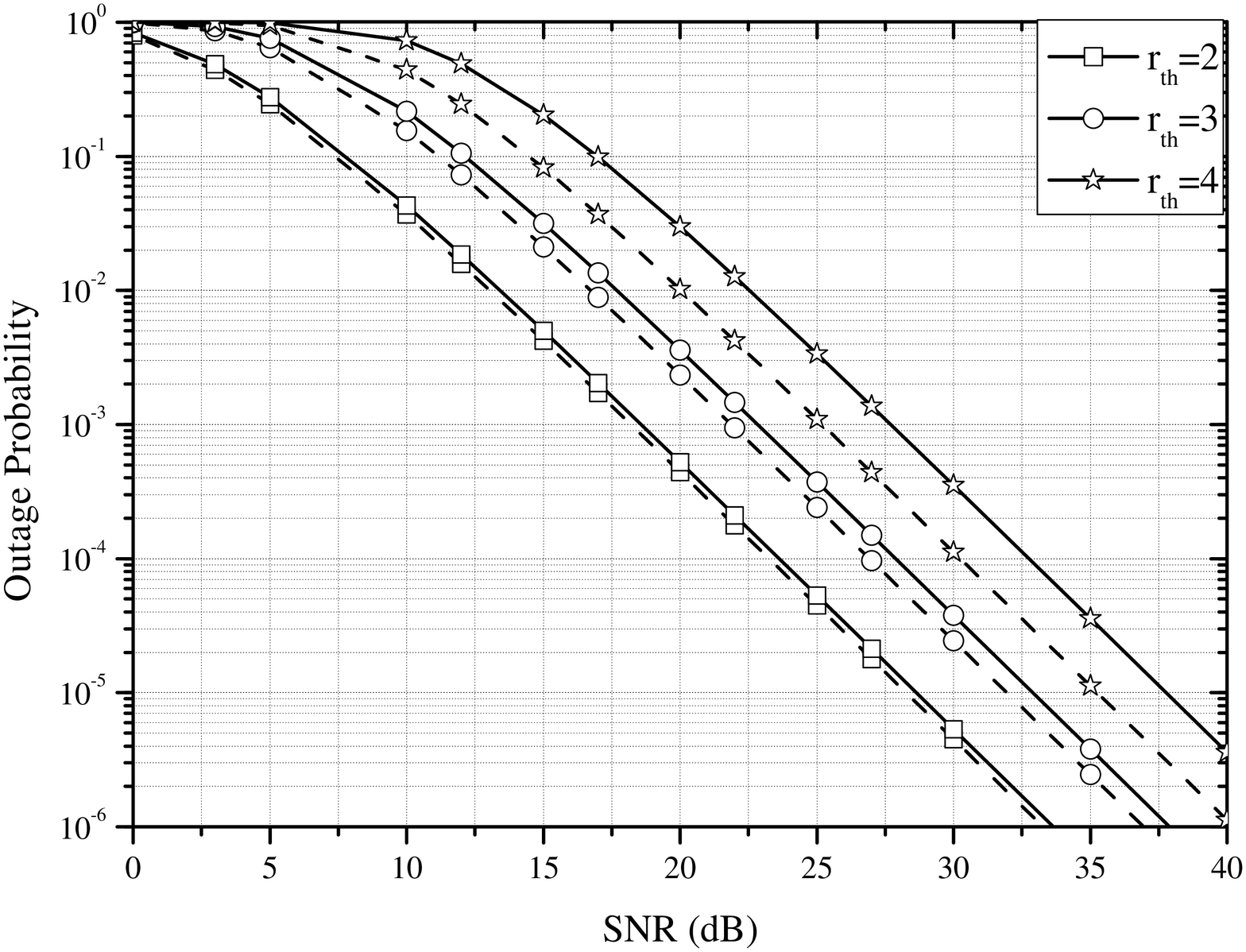}
\caption{Outage probability as a function of the SNR for different values of~$r_{\mathrm{th}}$, assuming ideal (dashed lines) and non-ideal (continious lines) RF chains with $N_t=2$, $k_{b,1}=0.08$, $k_{b,2}=0.17$, and $k_u=0.1$.}\label{fig:OP_vs_rate}
\end{figure}

Fig.~\ref{fig:OP_vs_rate} depicts the outage probability as a function of SNR, for different values of $r_{\mathrm{th}}$, assuming $N_t=2$, $k_{b,1}=0.08$, $k_{b,2}=0.17$, and $k_u=0.1$ (continuous lines), as well as the ideal RF chain case (dashed lines). Proposition~\ref{prop1} show perfect agreement with the marker symbols, which are the Monte-Carlo simulation results. For this figure, we observe that, in the low rate threshold, there is only a minor performance loss cause by the hardware imperfections. On the other hand, when the threshold is increased to $r_{\mathrm{th}}=4$ bits/channel use, there is a substantial performance loss. In more detail, for $r_{\mathrm{th}}=2$  and $r_{\mathrm{th}}=4$ bits/channel use, the system experience losses less than $0.5$ dB and about $5$ dB in SNR, respectively. Additionally, we observe that the outage probability curves with non-ideal hardware have the same slop as the ones with the ideal RF chains; therefore, hardware imperfections result in an SNR~offset.  

\begin{figure}
\centering\includegraphics[width=0.9\linewidth,trim=0 0 0 0,clip=false]{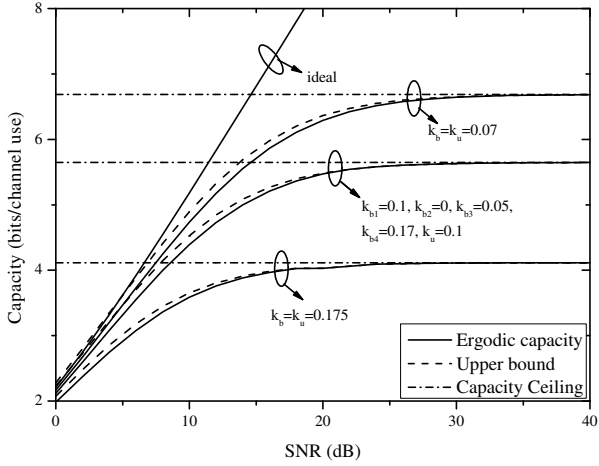}
\caption{Ergotic capacity as a function of the SNR for different levels of hardware imperfections and $N_t=4$.}\label{fig:Capacity}
\end{figure}

In Fig.~\ref{fig:Capacity}, the ergodic capacity is plotted as a function of the SNR, for different levels of imperfections. 
In this figure, the continuous lines represent Monte Carlo simulation results, while the dashed lines stand for the capacity upper bound, which is derived by~\eqref{Eq:C_bound} and~\eqref{Eq:CapacityBoundSpecialCase}. Additionally, the dashed-dotted lines denote the capacity ceiling, given by~\eqref{Eq:limit1}. Note that, as a benchmark, the capacity versus SNR for the ideal RF chains case is also plotted. 
As expected, as SNR increases, the capacity also increases.
However, in the case of non-ideal RF chains, the ergodic capacity saturates as approaches the capacity ceiling, as proven from Corollary~\ref{CorollaryCeiling}. 
As a result, since the capacity ceiling is determined by the level of imperfections, it increases as  $k_{b,i}$, ($i=1,\cdots, N_{t}$) and $k_u$ decreases. 
Moreover, we observe that the hardware imperfections have a small effect on the ergodic capacity at the low SNR regime, whereas, at the high SNR regime their impact is detrimental. 
Finally, from this figure, it is evident that the upper bound derived by Proposition~\ref{prop2} and~\eqref{Eq:CapacityBoundSpecialCase}, can be used as a simplified capacity approximation. This approximation can be considered tight in the high SNR regime.   

\begin{figure}
\centering\includegraphics[width=0.9\linewidth,trim=0 0 0 0,clip=false]{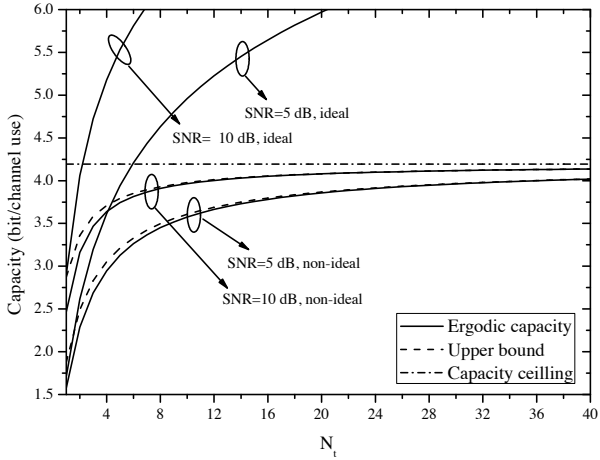}
\caption{Ergotic capacity as a function of the $N_t$ for different levels of SNR and $k_{b,i}=k_{u}=0.17$, $i\in[1, N_t]$.}\label{fig:Capacity2}
\end{figure}

In Fig.~\ref{fig:Capacity2}, the ergodic capacity is illustrated as a function of the number of antennas for SNR equals $5$ and $10$ dB and $k_{b,i}=k_u=0.17$ with $i\in[1, N_t]$. Note that the use of a high ammount of $N_t$ is a realistic scenario in mmW and THz communications. Moreover, since, in digital beamforming, as the number of antennas increases, the number of corresponding RF chains also increases; hence, the power consumption increases.  
In this figure, the continuous lines represent Monte Carlo simulation results,  the dashed lines stand for the capacity upper bound, which is derived by\eqref{Eq:CapacityBoundSpecialCase}, and the dashed-dotted lines denote the capacity ceiling.
From this figure, we observe that in the case of non-ideal RF chains, for a given SNR values, as the number of antennas increases, the ergodic capacity saturates and approaches $\log_{2}\left(1+\frac{1}{k_{b}^2+k_u^2}\right)$. 
Interestingly, the maximum achievable ergodic capacity is independent of the number of antennas at the BS, but it depends on the level of imperfections at both the BS and UE RF chains. 
This indicates that the key parameters in order to select the operational number of transmit antennas are the values of $k_{b}$ and $k_u$. 

\section{Conclusions}\label{S:Conclusions}
In this paper, we quantified the impact of hardware imperfections in wireless beamforming systems. 
In more detail, we provided simple closed form expressions for the system's outage probability and upper bounds for the ergodic capacity. 
These expressions take into account the level of imperfections at both the BS and the UE, the transmission and noise power, as well as the number of antennas at the BS. 
Moreover, through simulation results, we observed that the derived ergodic capacity upper bound can be used as a tight capacity approximation in the high SNR regime. 
Our results also revealed that there is a capacity ceiling that is independent of the number of transmit antennas, but it is determined by the level of imperfections. 
Therefore, there is a specific number of transmit antennas after which any further increase will not result to a further significant capacity gain. 
In other words, it is important for the designer of a digital beamforming system to take into account the level of imperfections in order to select the appropriate number of transmit~antennas.  

\section*{Appendices}
\section*{Appendix A}
\section*{Proof of Proposition 1}
The outage probability is defined as~\cite{Probability_book_stark}
\begin{align}
P_o(r_{\mathrm{th}}) &= P_r\left(\log_2\left(1+\gamma\right)\leq r_{\mathrm{th}}\right)  = P_r\left(\gamma\leq \gamma_{\mathrm{th}}\right),
\label{Eq:OP_def}
\end{align}
where $\gamma_{\mathrm{th}}=2^{r_{\mathrm{th}}}-1$.

By substituting~\eqref{Eq:SDNR} into~\eqref{Eq:OP_def}, and after some algebraic manipulations,~\eqref{Eq:OP_def} can be equivalently written~as
\begin{align}
P_o(r_{\mathrm{th}}) &= P_{r}\left(\mathcal{X}\leq \frac{\gamma_{\mathrm{th}}}{\tilde{\gamma}}\right)=F_{\mathcal{X}}\left(\frac{\gamma_{\mathrm{th}}}{\tilde{\gamma}}\right), 
\end{align}
where $F_{\mathcal{X}}(\cdot)$ denote the cumulative distribution function (CDF) of $\mathcal{X}$, whereas
\begin{align}
\mathcal{X}=\sum_{i=1}^N\left(1-\left(k_{b,i}^2 + k_u^2\right)\gamma_{\mathrm{th}}\right) |h_{0i}|^2.
\label{Eq:X}
\end{align}

Since $\mathbf{h}$ follows a Saleh-Valenzuela distribution, $|h_{0,i}|$, $i\in[1, N]$, is a Rayleigh distribution random variable. As a consequence, $\mathcal{X}$ is a sum of squared Rayleigh random variable with CDF obtained as~\cite{A:Karagiannidis-2006-ID448}
\begin{align}
F_{\mathcal{X}}(x) = \sum_{i=1}^{N_t} \Xi(i,1, b_i) \left(1 - e^{-\frac{x}{b_i}}\right).
\label{Eq:CDF}
\end{align}
By setting $x=\frac{\gamma_{\mathrm{th}}}{\tilde{\gamma}}$ into~\eqref{Eq:CDF}, we obtain~\eqref{Eq:OP}. This concludes the proof.

\section*{Appendix B}
\section*{Proof of Proposition 2}
The ergodic capacity is defined~as
\begin{align}
C=\mathbb{E}\left[\log_2\left(1+\gamma\right)\right],
\end{align}
or equivalently
\begin{align}
C = \mathbb{E}\left[\log_2\left(1+\frac{\rho}{q}\right)\right],
\label{Eq:C_def}
\end{align}
where, according to~\eqref{Eq:SDNR}, $\rho$ and $q$ can be expressed~as
\begin{align}
\rho= \sum_{i=1}^{N_t} |h_{0i}|^2 P_s
\label{Eq:rho}
\end{align}
and
\begin{align}
q= \sum_{i=1}^{N_t} (k_{b,i}^2+k_u^2)|h_{0i}|^2 P_s + N_0.
\label{Eq:q}
\end{align}

By applying Jensen's inequality~\cite{B:Handbook_of_complex_variables} into~\eqref{Eq:C_def}, we~obtain
\begin{align}
C\leq \log_2\left(1+\frac{\mathcal{A}}{\mathcal{B}}\right),
\label{Eq:C_inquality}
\end{align}
where $\mathcal{A}=\mathbb{E}\left[\rho\right]$ and $\mathcal{B}=\mathbb{E}\left[q\right]$. Next, we evaluate $\mathcal{A}$ and $\mathcal{B}$.

By taking into consideration~\eqref{Eq:rho}, $\mathcal{A}$ can be rewritten~as
\begin{align}
\mathcal{A}=P_s \mathbb{E}\left[\sum_{i=1}^{N_t}|h_{0,i}|^2\right].
\end{align}
Since $|h_{0,i}|$ follows Rayleigh distribution $\sum_{i=1}^{N_t}|h_{0,i}|^2$ is a chi-square distributed random variable with $N_t$ degrees of freedom. Therefore, $\mathbb{E}\left[\sum_{i=1}^{N_t}|h_{0,i}|^2\right]=N_t$ and 
\begin{align}
\mathcal{A}= P_s N_t.
\label{Eq:A}
\end{align}

From~\eqref{Eq:X} and~\eqref{Eq:q}, $\mathcal{B}$ can be expressed~as
\begin{align}
\mathcal{B}=P_s \mathbb{E}[\mathcal{Y}] + N_0,
\label{Eq:B}
\end{align}
where
\begin{align}
\mathcal{Y} = \sum_{i=1}^N\left(k_{b,i}^2 + k_u^2\right) |h_{0i}|^2.
\label{Eq:Y}
\end{align}
and  the expected value of $\mathcal{X}$ can be evaluated~as
\begin{align}
\mathbb{E}[\mathcal{Y}] = \int_{0}^{\infty} x f_{\mathcal{X}}(x) \mathrm{d}x,
\label{Eq:E_X}
\end{align}
where $f_{\mathcal{Y}}(x)$ is the probability density function of $\mathcal{Y}$. 
Note that  $\mathcal{Y}$, similarly to $\mathcal{X}$, follows sum of square Rayleigh distribution. Therefore, its CDF, $F_{\mathcal{Y}}(x)$ can be obtained by replacing $b_i$ with $c_i$ into~\eqref{Eq:CDF}, where the probability density function (PDF)  can be evaluated~as
\begin{align}
f_{\mathcal{Y}}(x) = \frac{\mathrm{d}F_{\mathcal{Y}}(x)}{\mathrm{d}x} = \sum_{i=1}^{N_t}\frac{\Xi(i, 1, c_i)}{c_i} e^{-\frac{x}{c_i}}. 
\label{Eq:PDF}
\end{align}
By substituting~\eqref{Eq:PDF} into~\eqref{Eq:E_X}, and carrying out the integration, we get
\begin{align}
\mathbb{E}[\mathcal{Y}] = \sum_{i=1}^{N_t}{\Xi(i,1, c_i)}{c_i}.
\label{Eq:E_X_final}
\end{align} 
By substituting~\eqref{Eq:E_X_final} into~\eqref{Eq:B}, we obtain
\begin{align}
\mathcal{B}=P_s \sum_{i=1}^{N_t}{\Xi(i,1, c_i)}{c_i} + N_0.
\label{Eq:B_final}
\end{align}
Finally, by substituting~\eqref{Eq:A} and~\eqref{Eq:B_final} into~\eqref{Eq:C_inquality}, we get~\eqref{Eq:C_bound}.  This concludes the proof.

\section*{Acknowledgment}
This work has received funding from the European Commission's Horizon 2020 research and innovation programme under grant agreement No 761794.
\balance
\bibliographystyle{IEEEtran}
\bibliography{IEEEabrv,References}

\end{document}